\documentclass[%
 reprint,
 amsmath,amssymb,
 aps,
]{revtex4-2}

\usepackage{xcolor}
\usepackage{graphicx}
\usepackage{dcolumn}
\usepackage{bm}
\usepackage{hyperref}
\usepackage{lineno}
\usepackage{xspace}

\newcommand{\kT}{$k_T$\xspace}
\newcommand{\akT}{anti-$k_T$\xspace}
\newcommand{\pT}{$p_T$\xspace}
\newcommand{\pp}{$p$+$p$\xspace}
\newcommand{\Au}{Au+Au\xspace}
\newcommand{\Pb}{Pb+Pb\xspace}
\newcommand{\GeV}{GeV/$c$\xspace}
\newcommand{\sNN}{$\sqrt{s_{\mathrm{NN}}}$\xspace}
\newcommand{\sqrts}{$\sqrt{s}$\xspace}

\newcommand{\fref}[1]{Fig.~\ref{#1}}
\newcommand{\eref}[1]{eq.~\ref{#1}}

\newcommand{\Fref}[1]{Figure~\ref{#1}}

\newcommand{\pikp}{$\pi^{\pm}$, K$^{\pm}$, p and $\bar{p}$\xspace}
\newcommand{\BG}{TennGen\xspace}
\newcommand{\PYTHIA}{\textsc{PYTHIA}\xspace}

\begin{document}

% \preprint{APS/123-QED}

\title{Multiplicity Based Background Subtraction \\ for Jets in Heavy Ion Collisions}% Force line breaks with \\
\author{Tanner Mengel} 
\author{Patrick Steffanic} 
\author{Charles Hughes}
\author{Antonio Carlos Oliveira Da Silva} 
\author{Christine Nattrass} 
\affiliation{University of Tennessee, Knoxville, TN, USA-37996.}

\date{\today}% It is always \today, today,

\begin{abstract}

Jet measurements in heavy ion collisions at low jet momentum can provide constraints on the properties of the quark gluon plasma 
but are overwhelmed by a significant, fluctuating background. We build upon our previous work which demonstrated the ability of 
the jet multiplicity method to extend jet measurements into the domain of low jet momentum~\cite{Mengel:2023mnw}. We extend 
this method to a wide range of jet resolution parameters. We investigate the over-complexity of non-interpretable machine 
learning used to tackle the problem of jet background subtraction through network optimization. Finally, we show that the 
resulting shallow neural network is able to learn the underlying relationship between jet multiplicity and background fluctuations, 
with a lesser complexity, reinforcing the utility of interpretable methods.

\end{abstract}

\maketitle

%%%%%%%%%%%%%%%%%%%%%%%%%%%%%%%%%%%%%%%%%%%%%%%%%%%%%%%%%%%%%%%%%%%%%%%%%%%%
% Introduction
%%%%%%%%%%%%%%%%%%%%%%%%%%%%%%%%%%%%%%%%%%%%%%%%%%%%%%%%%%%%%%%%%%%%%%%%%%%%
\section{Introduction}

The Quark Gluon Plasma (QGP), a dense and hot liquid consisting of quarks and gluons, is created in high-energy heavy ion collisions 
at both the Relativistic Heavy Ion Collider (RHIC)~\cite{Adcox:2004mh,Adams:2005dq,Back:2004je,Arsene:2004fa} and the 
Large Hadron Collider (LHC)~\cite{ATLAS:2014ipv,CMS:2012bms,ALICE:2010yje}. QGP properties can be constrained by quantitatively comparing 
jet measurements of jets created by hard scatterings between partons with theoretical models within these collisions~\cite{Connors:2017ptx,Burke:2013yra,JETSCAPE:2021ehl}. 
Particles produced in heavy ion collisions are predominantly soft particles unrelated to hard interactions. The details of 
this background in jet measurements and its fluctuations are influenced by correlations arising from hydrodynamic flow and 
the shape of single particle spectra~\cite{Voloshin:2008dg}; thus, they are unlikely to perfectly match between empirical data and models. 
A background determined from  mixed events is consistent with the measured background in hadron-jet correlations measured by STAR~\cite{STAR:2013thw} 
at RHIC. Measurements of the width of the distribution of momentum in random cones by the ALICE Collaboration are consistent with expectations from 
particles drawn randomly from a Gamma distribution~\cite{ALICE:2012nbx}. This indicates that the background is largely consistent with a random selection 
of particles drawn from a momentum distribution , with some contribution from correlations due to flow~\cite{Hughes:2020lmo}.

The precision of jet measurements and their kinematic range are restricted by corrections for this background and its fluctuations.  Background fluctuations 
lead to a larger uncertainty in each individual jet's momentum.  Because the precise number of combinatorial jets is not known well, 
measurements are typically restricted to a kinematic region where their contribution is negligible.  Background fluctuations lead to smearing 
with a standard deviation independent of jet momentum and the combinatorial jet contribution is dominant at low momentum.  These effects therefore 
restrict the measurement of low momentum jets.  Extending the kinematic range of jet measurements to lower momenta would improve constraints on 
partonic energy loss.  In particular, since a greater fraction of gluon-like jets are at low momenta, this would increase sensitivity to modifications of gluon-like jets. 

Jet spectra measurements extending to the lowest momenta primarily use the area method~\cite{Soyez:2009cw} approach for background subtraction. This technique was 
initially formulated to mitigate the underlying event in \pp collisions under conditions of high pile-up~\cite{Soyez:2009cw}, and 
has been subsequently adapted for application in heavy ion collisions as well~\cite{ALICE:2013dpt,ALICE:2015mjv,ALICE:2019qyj,STAR:2020xiv}. 
The area method is usually used instead of iterative background subtraction methods~\cite{Hanks:2012wv,CMS:2016uxf,ATLAS:2012tjt} for 
measurements of jets at lower momenta. Iterative methods may suppress the background fluctuations by estimating the local background and 
suppress combinatorial jets by requiring high momentum or energy constituents.  At low momenta, these requirements may impose a bias on 
the surviving jets.  Fluctuations and the contribution from combinatorial jets are generally higher with the area method, but with less bias.

In nuclear physics, there has been a deliberate emphasis on advancing the utility of machine learning analysis techniques, with 
a specific focus on employing methods that are transparent, resilient, and capable of providing unambiguous quantification of 
uncertainties, all while remaining explainable~\cite{Achenbach:2023pba}.
Machine learning has been applied to subtract the background in jet measurements in heavy ion collisions~\cite{Haake:2018hqn,ALICE:2023waz}. 
Using machine learning techniques for background subtraction requires cautious handling, as models for the background fall short in 
replicating background fluctuations in heavy ion collisions~\cite{Hughes:2020lmo}. The utility of non-interpretable machine learning 
methods is limited in scenarios where the available training models may lack accuracy, where comprehension of the method is 
necessary to decipher outcomes, or when results are sought beyond the boundaries of the training data set. 
Moreover, the application of non-interpretable machine learning techniques that exhibit advancements over traditional background approaches often 
yields little to no insight into the underlying physical relationships harnessed to attain such enhancements. Using a deep neural network, 
characterized by multiple concealed layers, for jet background subtraction leading to considerably better resolution than the area method, 
particularly at low momenta~\cite{Haake:2018hqn,ALICE:2023waz}. However, deep neural networks are susceptible to model bias, as their predictions 
tend to be unreliable beyond the scope of their training dataset, and the efficacy of these methods may deteriorate when extended beyond this scope.  
Due to their opacity, they provide limited insight into the exact locations and causes behind such breakdowns. Less complex networks can provide 
clearer interpretations on the training bias and limits of applicability. 

The efficacy of machine learning techniques compared to traditional approaches indicates the presence of discernible information that these machine 
learning methods leverage for their improvements. In~\cite{Mengel:2023mnw}, we introduced an interpretable machine learning approach that allowed 
us to discern the causes of the improved momentum resolution seen when a deep neural network is used for background subtraction. We devised an 
alternative, physics-based technique, referred to as the ``multiplicity method”, which built upon the background outlined in~\cite{TANNENBAUM200129,ALICE:2012nbx}. 
By evaluating the widths of the jet momentum fluctuations using the multiplicity method and contrasting them with those of the area and neural network methods, 
we gauged the impact of these methodologies on the achievable kinematic range. We demonstrated with  symbolic regression that the neural network was learning 
an algorithm approximately the same as the multiplicity method.  

We expand on the work in~\cite{Mengel:2023mnw}.  We investigate methods for reducing the complexity of the neural network to demonstrate how 
comparable performance can be achieved with a simpler system.  We compare this shallow neural network to the deep neural network as well as the 
area and multiplicity methods.  We investigate the jet resolution parameter and collision energy dependence of these methods in greater detail. 
We also investigate the impact of unfolding for momentum resolution and the kinematic reach of these methods.

%%%%%%%%%%%%%%%%%%%%%%%%%%%%%%%%%%%%%%%%%%%%%%%%%%%%%%%%%%%%%%%%%%%%%%%%%%%%
% Method
%%%%%%%%%%%%%%%%%%%%%%%%%%%%%%%%%%%%%%%%%%%%%%%%%%%%%%%%%%%%%%%%%%%%%%%%%%%%
\section{Method}

\subsection{Simulation}

The \BG~\cite{Hughes:2020lmo,TennGen,TennGen200} model is designed to simulate heavy ion collisions by randomly producing \pikp hadrons.  
A random momentum is chosen from the momentum distribution observed in the data~\cite{PHENIX:2013kod,Abelev:2013vea}.  
For each event, even order event planes are fixed to $\phi=0$ and a random orientation is chosen for each odd order event plane.  
The expected azimuthal distribution for each particle is constructed and a random azimuthal angle is chosen from that distribution~\cite{PHENIX:2014uik,Adam:2016nfo}. 
It accurately reproduces the yields~\cite{Aamodt:1313050}, momentum distributions~\cite{PHENIX:2013kod,Abelev:2013vea}, and azimuthal 
anisotropies~\cite{PHENIX:2014uik,Adam:2016nfo} observed in published data. \BG was updated to incorporate collision energies per nucleon 
of \sNN = 200 GeV and \sNN = 2.76 TeV, as well as include multiplicity fluctuations~\cite{Mengel:2023mnw}.  The computational efficiency was improved as well.  
By design, \BG does not include correlations other than flow.  This is motivated by observations that the background for jet measurements is roughly 
consistent with that expected from particles randomly selected from the single particle momentum distribution~\cite{ALICE:2012nbx,Hughes:2020lmo,STAR:2013thw}. 
 
\PYTHIA 8.307~\cite{Sjostrand:2007gs} $pp$ events combined with \BG~\cite{TennGen,TennGen200} heavy ion backgrounds are used to model the environment 
found in heavy ion collisions.  The Monash 2013 tune~\cite{Skands:2014pea} was used to simulate proton-proton collisions at \sqrts = 200 GeV and 2.76 TeV. 
The simulations were conducted in 25 $p_{T}^{\text{hard}}$ bins, starting at 10 GeV, and each bin contained 1 million \pp events. In this simulation, only final 
state charged particles from PYTHIA were merged with a \BG background event. Both \PYTHIA and \BG particles are restricted to $p_T>150$ MeV/$c$ and pseudorapidity $|\eta| <$ 0.9.

Jet finding is done on the combined event using the \akT algorithm with the standard recombination scheme implemented with FastJet 3.4.0~\cite{Cacciari:2011ma}. 
The jet resolution parameters, denoted as $R$, are set to values of 0.2, 0.4, and 0.6. The \PYTHIA event is initially clustered separately prior to merging for 
all jet resolution parameters, which we use to define the truth jet momentum. After merging the combined event is re-clustered and the jets in the combined 
\PYTHIA and \BG event are geometrically matched to a \PYTHIA jet if $\Delta R = \sqrt{\Delta\eta^{2} + \Delta\phi^{2}} < 0.1$ where $\Delta\eta$ and $\Delta\phi$ 
are the differences in $\eta$ and $\phi$ between the jets and there is a bijective match. The momentum of the matched \PYTHIA jet is taken as the true momentum for 
the combined jet, denoted by $p_{T,\text{jet}}^{\text{Truth}}$, which is equivalent to $p_{T,\text{jet}}^{\text{PYTHIA}}$. Kinematic cuts on jet momentum 
$p_{T,\text{jet}}^{\text{raw}} > 10 \text{ GeV}$ and pseudorapidity $|\eta_{\text{jet}}| < 0.9 - R$ are imposed on the final reconstructed jets from the combined event. 

\subsection{Area method}

The area method, introduced by in~\cite{Cacciari:2008gn}, corrects jet momentum by estimating the average energy density of the background and 
assuming its uniform distribution across the jet area. The corrected jet momentum is
\begin{equation}
    p_{T,\text{jet}}^{\text{Corr, A}} = p_{T,\text{jet}}^{\text{raw}} - \rho A , \label{Eq:pTArea}
\end{equation}
where the jet area $A$ is computed utilizing ``ghost" particles and $\rho$ is the background momentum density per unit area. For a given event, 
$\rho$ is approximated as the median of $p_{T,\text{jet}}^{\text{tot}}/A$ for \kT jets, primarily due to the dominance of background within \kT jets.

The standard deviation of the momentum residual for the area method is approximately the same as the standard deviation of the momentum in a 
random cone~\cite{ALICE:2012nbx,Hughes:2020lmo,TANNENBAUM200129}
\begin{equation}
    \sigma_{\delta p_{T}^{\text{Cone}}} = \sqrt{N \sigma^{2}_{p_T} + (N + 2 N^2 \sum_{n=1}^{\infty} v_n^2) \langle p_{T} \rangle^{2}} , \label{eq:DeltaPtwidths_flow}
\end{equation}
where $N$ is the number of background particles, $\sigma_{p_{T}}$ is the standard deviation of the momentum of single particles, $v_n$ denotes the 
coefficients of azimuthal anisotropies of single particles, and $\langle p_T \rangle $ is the average momentum of background particles~\cite{ALICE:2012nbx}. 
This derivation assumes that each of the $N$ particles is drawn from a single track momentum distribution, which can be approximated as a Gamma distribution 
accounting for the first term~\cite{TANNENBAUM200129}. The second term is attributed to Poissonian fluctuations in the number of background particles, while the 
third term arises from fluctuations in particle count due to hydrodynamical flow. Slight variations in the width are observed due to deviations in the single 
track momentum distribution from a Gamma distribution and the momentum-dependent nature of $v_n$~\cite{Hughes:2020lmo}.

\subsection{Multiplicity method}

We proposed an alternative method in~\cite{Mengel:2023mnw}, the multiplicity method, as a substitute for the area method. In this approach, we 
use the average momentum of background particles $\langle p_T \rangle $ and the average excess multiplicity originating from background within 
the jet to subtract the average background energy
\begin{equation}
    p_{T,\text{jet}}^{\text{Corr, N}} = p_{T,\text{jet}}^{\text{raw}} - \rho_{\text{Mult}}(N_{\text{tot}}- \langle N_{\text{signal}} \rangle) ,\label{Eq:pTN}
\end{equation}
where $N_{\text{tot}}$ is the observed number of particles within the jet, $\langle N_{\text{signal}} \rangle$ is the average number of particles in the signal, 
and $N = N_{\text{tot}} - \langle N_{\text{signal}}\rangle $. Additionally, $\rho_{\text{Mult}}$ within an event represents the mean transverse momentum 
per background particle. Similar to the area method $\rho_{\text{Mult}}$ is approximated as the median of $p_{T,\text{jet}}^{\text{tot}}/N_{\text{tot}}$ for 
\kT jets. This approach capitalizes on the fact that the natural variable in the standard deviation is the number of background particles, eliminating the 
second and third terms in equation~\ref{eq:DeltaPtwidths_flow}.  The standard deviations of the momentum residual for the multiplicity method is therefore approximately
\begin{equation}
    \sigma_{\delta p_{T}} = \sqrt{N} \sigma_{p_{T}}.
\end{equation}
There will still be some influence from deviations in the shape of the single particle spectrum and a more accurate accounting of flow, but contributions 
from number fluctuations are eliminated.

\begin{figure*}
    \centering
    \includegraphics[width=2\columnwidth]{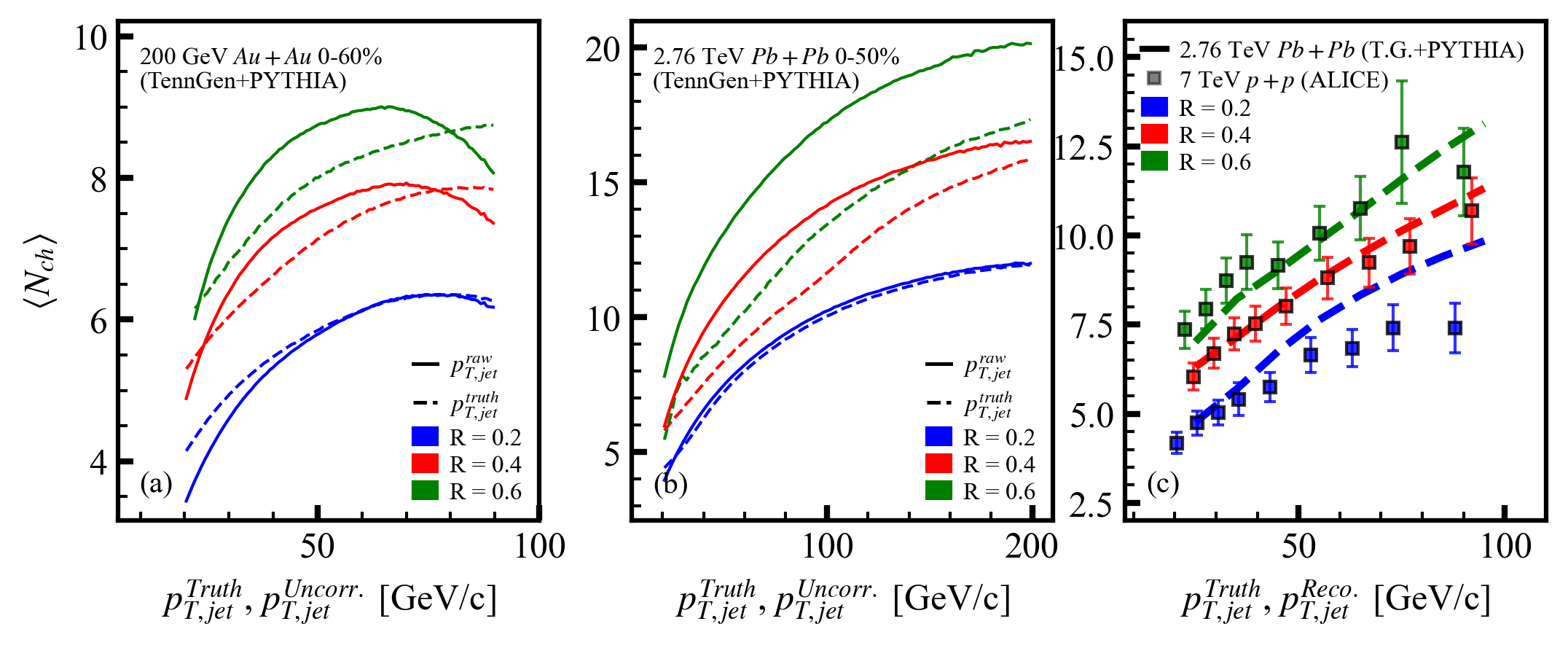}
    \caption{Estimations for $\langle N_{signal} \rangle $ for (a) \Au collisions at \sNN = 200 GeV and (b) \Pb collisions at \sNN = 2.76 TeV for jet resolution parameters 
    R = 0.2, 0.4, and 0.6. (c) Comparisons between estimated  $\langle N_{signal}\rangle$ and measured jet multiplicity for \pp collisions at $\sqrt{s} =$ 7 TeV~\cite{ALICE:2014dla}.}\label{fig:NchargeEstimates}
\end{figure*}

The parameter $N_{\text{signal}}$ can be estimated with accuracy, as models for jets in proton-proton collisions~\cite{ALICE:2014dla} can 
adequately describe it. In our study, we use the average number of particles in a jet $\langle N_{\text{signal}} \rangle $ reconstructed 
with a given momentum. The average \PYTHIA jet multiplicity corresponding to a specific jet \pT bin is used to estimate $N_{\text{signal}}$ for a jet. 
The estimations of $N_{\text{signal}}$ for \Au and \Pb collisions at \sNN = 200 GeV and 2.76 TeV are displayed in Fig.~\ref{fig:NchargeEstimates}.

While this method introduces an additional systematic uncertainty because $\langle N_{\text{signal}} 
\rangle $ must be estimated with a model, this may still decrease the overall uncertainties because of the reduction in background fluctuations. 
Since the dependence of $\langle N_{\text{signal}} \rangle $ on $p_{T,\text{jet}}^{\text{Truth}}$ and $p_{T,\text{jet}}^{\text{raw}}$ is similar 
and PYTHIA agrees well with the data, a reasonable uncertainty on $\langle N_{\text{signal}} \rangle$ for unmodified jets is projected to be less than one for $R \leq 0.4$ and around two for $R=0.6$.

Additional constraints are required to account for medium modifications to jet fragmentation in heavy ion collisions. Jet multiplicity 
is largely dependant on jet momentum scaling therefore we estimate the effect in both the low and high jet momenta regime~\cite{Lee:2023tfx}. 
Measurements of jet fragmentation functions with $p_{T}^{\text{jet}}>100$ \GeV in \Pb collisions~\cite{CMS:2014jjt,ATLAS:2018bvp} suggest the presence 
of at most one additional particles.  Measurements of gamma-hadron correlations at RHIC~\cite{PHENIX:2020alr}, covering a much lower kinematic range, 
likewise indicate that there are fewer than one additional particles due to medium modifications.

The additional uncertainty in jet energy introduced by the multiplicity method is proportional to $\sigma_{N_{\text{signal}}}\cdot\rho_{\text{Mult}}$, 
where $\sigma_{N_{\text{signal}}}$ is the uncertainty in the count of signal particles in the jet. Given that $\rho_{\text{Mult}} \approx 0.5$ \GeV, 
this uncertainty would approximately amount to $1.5$ GeV for medium modifications and around 0.5--1.0 GeV for uncertainties in the count of signal 
particles in unmodified jets. These values are small compared to typical experimental jet momentum resolutions of 10-20${\%}$.

We estimate that the multiplicity method would be better than the area method when the extra terms in \eref{eq:DeltaPtwidths_flow} are larger 
than the uncertainty due to $\sigma_{N_{\text{signal}}}$.  Since $N \approx \frac{R^2}{2} \frac{dN_{ch}}{d\eta}$ for a random cone where 
$\frac{dN_{\text{ch}}}{d\eta}$ is the charged particle multiplicity in the event and $\rho_{Mult} \approx \langle p_T \rangle$, the multiplicity method is generally better for
\begin{equation}
    \frac{dN_{ch}}{d\eta} 	\gtrapprox \frac{2 \sigma_{N_{\text{signal}}}^2}{R^2} ,
\end{equation}
which corresponds to thresholds of $\frac{dN_{\text{ch}}}{d\eta} = 450, 113, $ and 50 for $R = 0.2, 0.4,$ and 0.6, respectively, 
for $\sigma_{N_{signal}} = 3$.  This indicates that gains from the multiplicity method may not be significant for $R=0.2$ jets at 
RHIC, where multiplicities only reach $dN_{ch}/d\eta \approx 687$~\cite{PHENIX:2015tbb} in 0--5\% central \Au collisions at \sNN = 200 GeV, 
but are likely significant for wider jets and at the LHC.

\subsection{Machine learning methods}\label{Sec:MachineLearning}

In this section we present an overview of machine learning methods used in this study. These methods include the deep neural network 
(DNN) inspired by~\cite{Haake:2018hqn}, our pruned shallow neural network (SNN) which demonstrates the dimensionality of the problem of 
jet background subtraction and finally, the deep symbolic regression method which allows for direct observation of the mapping learned by 
the neural networks. All neural networks are implemented in TensorFlow 2.10.0~\cite{tensorflow2015-whitepaper} and optimized using ADAM~\cite{GoodBengCour16}. 
The loss functions for all machine learning techniques is some form of the mean squared error between the target jet momenta and the predicted. Unless 
otherwise denoted the nodes of the neural networks are activated with a rectified linear unit (ReLU)~\cite{GoodBengCour16}. The population of simulated 
jets are randomly split 50/50 and used as training and testing datasets.

\subsubsection{Deep Neural Network Method}\label{Sec:DNNMethod}

Deep Neural Networks are a class of machine learning model that have achieved significant success in a wide range of tasks, including image 
classification, natural language processing, and feature regression. The deep neural network architecture used in this study is composed 
of multiple hidden layers of artificial neurons with trainable weights connecting each of the nodes of a layer to each of the neurons of the next. 
The relationship describing the propagation from layer to layer can be generalized as
\begin{equation}\label{Eqn:Loss}
    \psi_{i} = \mathcal{A}(\mathbf{W}_{i}\cdot\psi_{i-1}+b_{i}),
\end{equation}
\noindent where $\psi_{i}$ is an n-dimensional space representation of $\psi_{i-1}$, $\mathbf{W}_{i}$ is an $n\times m$ dimensional weight matrix which is 
applied to the m-tuple $\psi_{i-1}$, $b_{i}$ is the learned bias, and $\mathcal{A}$ is the activation function. The activation function of a 
neuron is typically a non-linear function which sets the threshold for a neuron to fire, i.e propagate its input onto the next hidden layer. 
Through the non-linearity imposed by activation functions, deep neural networks are capable of learning complex relationships in data by 
optimizing trainable weights to minimize a loss function. 

We train our deep neural network to predict the corrected jet momentum from the following input features: the uncorrected jet momentum, 
jet area, jet angularity, number of jet constituents, seven leading constituent transverse momenta. The architecture and input features of 
the network are motivated by previous application of neural networks to proton-proton jets with a thermal background~\cite{Haake:2018hqn}. 
The deep neural network has three hidden layers consisting of 100, 100 and 50 nodes, and the loss function is the mean squared error
\begin{equation}\label{Eqn:loss}
  \mathcal{L} =  \frac{1}{n_{\text{jets}}}\sum_{i=1}^{n_{\text{jets}}}(p_{T,\text{jet}_i}^{\text{Truth}} - p_{T,\text{jet}_i}^{\text{DNN}})^2,
\end{equation}
where $p_{T,\text{jet}}^{\text{DNN}}$ is the predicted jet momentum, $p_{T,\text{jet}}^{\text{Truth}}$ is the truth momentum. 

\subsubsection{Shallow Neural Network Method}\label{Sec:SNNMethod}

\begin{figure}
    \centering
    \includegraphics[width=\columnwidth]{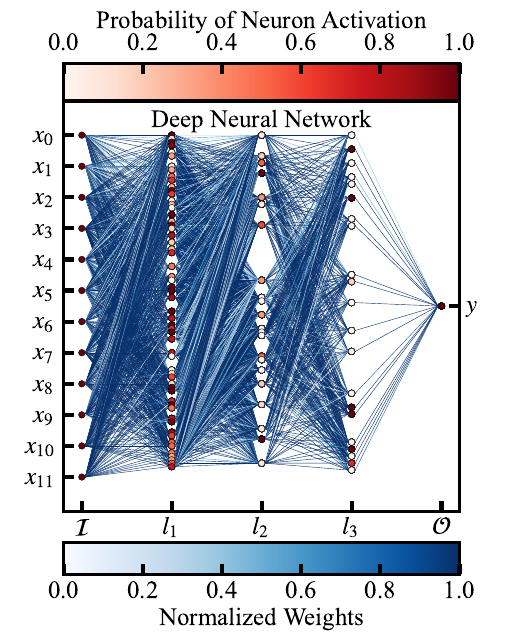}
    \caption{Each node and weight for every layer $l$ in the deep neural network architecture. The nodes are colored by their probability of 
    firing for the entire testing jet sample. The weights are normalized to the maximum weight from a given node to the next layer. The input 
    features are listed by $x_i$ where $i$ ranges from $0$ to $11$ corresponding to the 12 input features for the deep neural network 
    (uncorrected jet momentum, jet area, jet angularity, number of jet constituents, and seven leading constituent momenta). The truth jet momentum is labeled by y.} 
    \label{fig:dnnSparse}
\end{figure}

To demonstrate the complexity of the deep neural network architecture presented in~\cite{Haake:2018hqn}, the fully trained deep neural 
network is shown in \fref{fig:dnnSparse}. Even prior to pruning the network fails to use all connections available to it. While 
this network offers improved jet momentum resolution, its over-complex architecture lacks interpretability, hindering our ability to 
extract meaningful insights from the learned representations~\cite{GoodBengCour16}.

To highlight the important connections formed by the deep neural network we introduce network pruning to reduce the complexity of deep neural 
network and eliminate redundant parameters/connections that are deemed unnecessary for achieving optimal performance~\cite{li2017pruning}. 
The pruning procedure involves training a network architecture, assessing each of the nodes probability of firing over the entire testing dataset 
and then removing those nodes which do not reach the threshold of firing. In this study our threshold was set to zero, meaning only unused nodes 
where removed. After pruning unused nodes, the network is retrained and then the loss is compared to the original loss of the un-pruned network. If a 
layer drops below 3 used nodes, the remaining nodes are added to the previous layer and that layer is removed from the architecture.  This process 
is repeated until the loss grows beyond 5${\%}$ of the original loss. Additional optimization was done on the shallow network through input feature engineering. 
The best input features through multiple statistical tests where determined to be the jet area, the multiplicity, the uncorrected jet momentum, the 
angularity and the leading hadron momentum. The details of these metrics are outline in the appendix.

\begin{figure}
    \centering
    \includegraphics[width=\columnwidth]{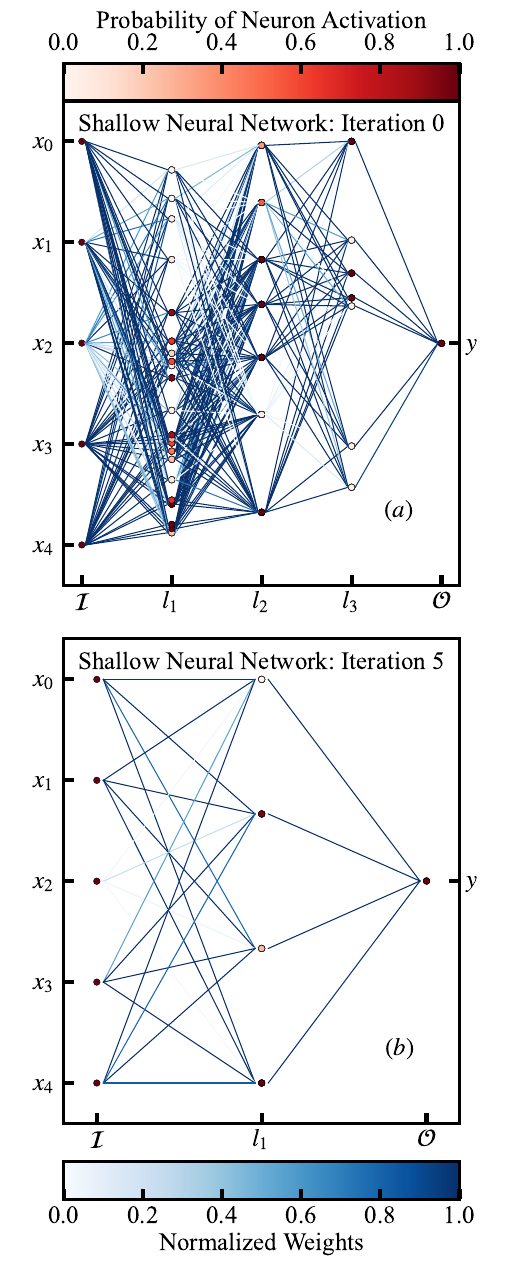}
    \caption{Each node and weight for every layer $l$ in for the first (a) and last (b) iteration of the shallow neural network 
    produced through feature optimization and network pruning. The nodes are colored by their probability of firing for the entire 
    testing jet sample. The weights are normalized to the maximum weight from a given node to the next layer. The input features are 
    listed by $x_i$ where $i$ ranges from $0$ to $4$ corresponding to uncorrected jet momentum, leading constituent momenta, number of jet 
    constituents,jet area, and jet angularity. The truth jet momentum is labeled by y.}\label{fig:snnIterations}
\end{figure}

During the pruning procedure, kernel regularization is employed to further simplify the network by discouraging unnecessary the weight parameters. 
Kernel regularization is a technique that imposes a penalty on the weight parameters of a neural network by adding a regularization term,
\begin{equation}\label{Eqn:DNNloss}
  \mathcal{L}' = \mathcal{L}_o + \lambda\sum_{l=1}^{L}||\mathbf{W}_{l}||^{2},
\end{equation}
\noindent to the original loss function $\mathcal{L}_o$, which is typically defined as the $L2$ norm of the weights $||\mathbf{W}_{l}||^2$ of layer $l$ 
summed over the $L$ layers and multiplied by a regularization parameter, $\lambda$. By minimizing the combined loss function comprising the original 
task-specific loss and the regularization term, the network is incentivized to prune out less important connections and nodes, leading to a more 
compact and interpretable architecture~\cite{lasso}. \Fref{fig:snnIterations} shows the initial and final iteration of this network pruning 
procedure with $\lambda = 0.01$. The resulting shallow neural network consists of just $29$ learned parameters, compared to the $16,501$ 
available to the deep neural network. 

\subsubsection{Deep symbolic regression}

\begin{figure*}
    \centering
    \includegraphics[width=2\columnwidth]{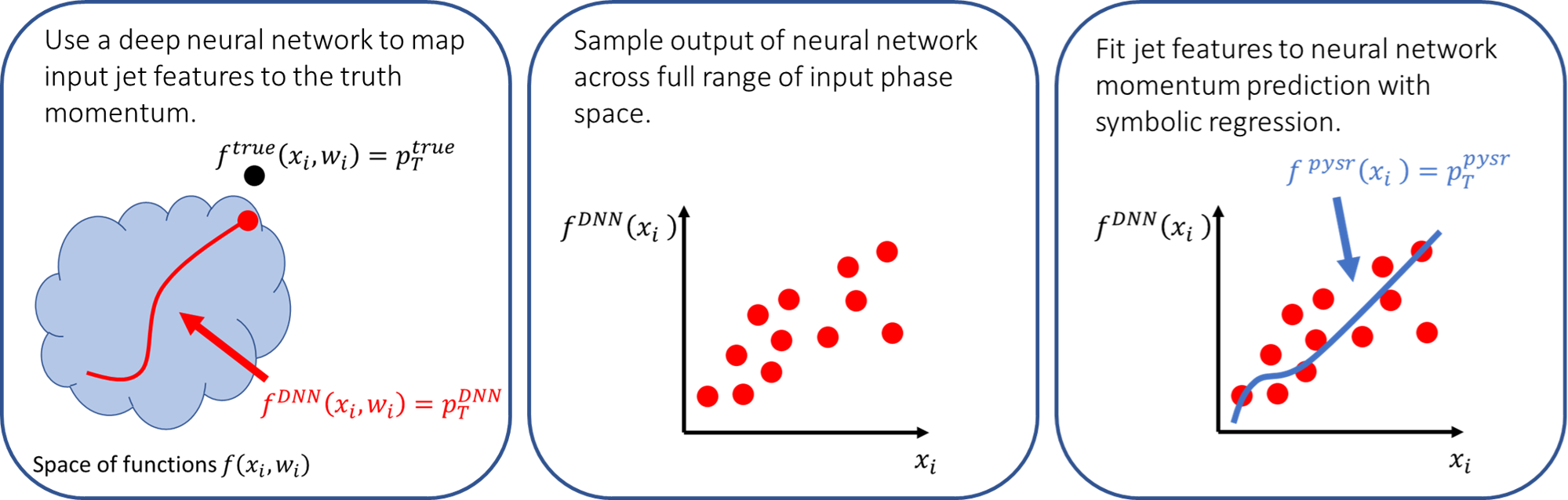}
    \caption{Diagram showing the procedure of mapping a relationship between input jet features to the corrected jet momentum 
    using a neural network and then extracting an analytical expression which describes that mapping with symbolic regression. 
    The left panel shows fitting the space of possible functions which map input to output with the neural network. The middle 
    panel shows sampling of this mapping across input jets. The right panel shows fitting this sample to an analytical expression using symbolic regression.}
    \label{fig:PySrDiagram}
\end{figure*}
   
A sufficiently complex neural network can interpolate any function, at the cost of transparency to the user. This poses an 
obstacle to application of deep neural networks in physics where understanding predictions and identifying their potential biases is crucial. 
Our approach to addressing this challenge is through symbolic regression, one example of interpretable machine learning, to extract mathematical 
expressions from trained deep neural networks. The resulting equations provide an effective description of the neural network's mapping between the 
input and output. By constraining the types of operations available, we can impose complexity and smoothness requirements. The process of extracting a 
functional representation from a trained deep neural network is shown schematically in figure~\ref{fig:PySrDiagram}.

Once a neural network is trained, it represents an approximate mapping between the input jet features and the truth jet momentum.
The training of the deep neural network outlined in sec.~\ref{Sec:DNNMethod}. Kernel regularization is employed to strengthen the learned mapping. The loss function 
is given by~\ref{Eqn:DNNloss} where regularization with $\lambda = 0.001$.

After the network is trained, we apply symbolical regression to extract a functional form which describes this mapping using the 
PySR 0.11.11~\cite{pysr} package. The PySR model samples the phase space of analytic expressions defined by operators, input features, 
and constants for minimization through genetic programming. The input features are identical to those of the deep neural network, and the pool of 
operations are arithmetic, exponential, trigonometric, and exponentiation. The model mutates over 50 generations of 20 different population samples, 
with each population containing 33 individuals. The loss function for the PySR model
\begin{equation}\label{Eqn:Pysrloss}
  \mathcal{L} = \frac{1}{n_{jets}}\sum_{i=1}^{n_{jets}}(p_{T,jet_i}^{DNN} - p_{T,jet_i}^{PySR})^2,
\end{equation}
is the mean squared error between the prediction from PySR momentum prediction $p_{T,jet}^{PySR}$ and the corrected jet momentum 
predicted by the deep neural network $p_{T,jet}^{DNN}$. PySR evaluates expressions based on a score $S$ that rewards minimizing the 
loss function $\mathcal{L}$ and penalizes equation complexity $C$
\begin{equation}\label{Eqn:Score}
  S = -\frac{d \ln\mathcal{L}}{d C}, 
\end{equation}
where the equation complexity $C$ is defined as the total number of operations, variables, and constants used in an equation~\cite{cranmer2020discovering}. 
The highest scoring PySR expression is a functional representation of the mapping from input jet features to corrected jet momentum learned by the deep neural network. 

In addition to mapping the deep neural network, the shallow neural network described in sec.~\ref{Sec:SNNMethod} was mapped using the same procedure. 

\subsection{Unfolding}\label{Sec:Unfolding}

The improvement in jet momentum resolution should extend the kinematic range of the measurement to lower jet momenta. This is tested by unfolding the reconstructed 
jet momentum spectra using five iterations of the Bayesian unfolding method~\cite{DAGOSTINI1995487} in RooUnfold 2.0.0~\cite{Brenner:2019lmf}. Unfolding is a 
procedure used to correct for smearing due to finite resolution in the uncorrected measurement. Five iterations is when the change in $\chi^2$ between the 
unfolded and truth spectra becomes less than the uncertainties of the measured spectra. Anymore iterations after five would not change the unfolded measurement 
within uncertainties. The ability for each method to reconstruct these combinatorial jets in low momenta bins will determine the kinematic reach of the unfolded 
jet spectra. Increased precision in jet momentum allows for combinatorial jets to be reconstructed in jet momentum bins closer to zero.
We construct a response matrix using \PYTHIA jets (truth jets) matched to \PYTHIA+\BG jets (reconstructed jets). The momentum of the \PYTHIA jet is taken as the 
truth momentum, $p_{T,jet}^{Truth} \equiv p_{T,jet}^{PYTHIA}$.  We then unfold our reconstructed jet spectra. The reconstructed spectra has no matching 
criteria between the \PYTHIA+\BG jets and \PYTHIA jets. The lower threshold for unfolding is typically set to be five times the width of the jet momentum 
resolution $\sigma_{\delta p_{T}}$ to suppress effects of combinatorial jets on the unfolded results~\cite{ALICE:2013dpt,ALICE:2023waz}. 
We use reconstructed jet spectra including combinatorial background to investigate the sensitivity of the lower momentum threshold to combinatorial 
background, since most of the combinatorial jets populate this kinematic region. We find that the width of the resolution does not impact the uncertainties on the unfolded spectra.

%%%%%%%%%%%%%%%%%%%%%%%%%%%%%%%%%%%%%%%%%%%%%%%%%%%%%%%%%%%%%%%%%%%%%%%%%%%%
% Results
%%%%%%%%%%%%%%%%%%%%%%%%%%%%%%%%%%%%%%%%%%%%%%%%%%%%%%%%%%%%%%%%%%%%%%%%%%%%
\section{Results}

\begin{figure*}
  \centering
  \includegraphics[width=2.0\columnwidth]{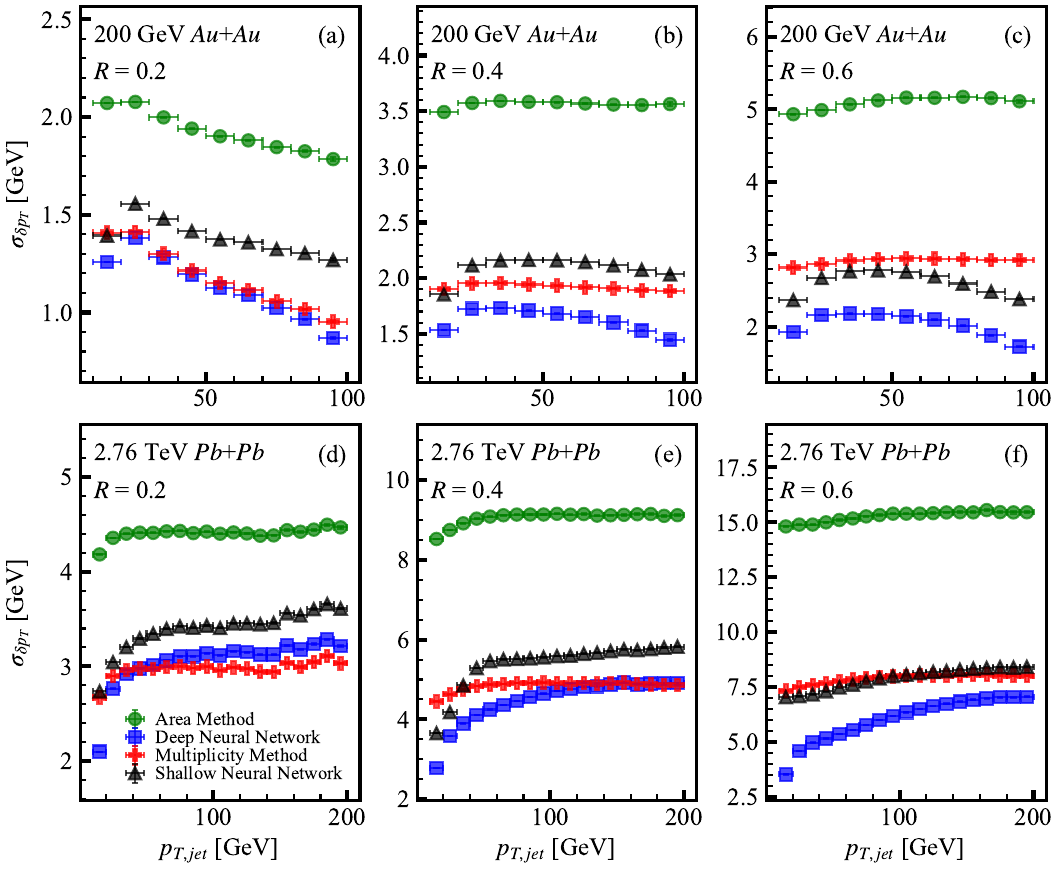}
  \caption{Comparisons of jet $p_{T}$ residual width for each background subtraction method as a function of reconstructed jet momentum 
  for \Au collisions at \sNN = 200 GeV and \Pb collisions at \sNN = 2.76 TeV for jet resolution parameters $R=0.2, 0.4,$ and $0.6$. Note 
  the $y-$axes are zero suppressed for all panels. 
  }
  \label{fig:ReconstructionWidthvspT}
\end{figure*}

In~\fref{fig:ReconstructionWidthvspT}, we observe the variations in the width of the jet momentum residual distributions versus the jet momentum 
for different background subtraction methods applied to jets with jet resolution parameters $R = 0.2$, $0.4$, and $0.6$ in both \Au collisions at \sNN = 200 
GeV and \Pb collisions at \sNN = 2.76 TeV. As expected, the width of the distribution, represented by $\sigma_{\delta p_{T}}$, increases as the jet resolution 
parameter grows, indicating a larger background contribution for larger jets. Additionally, the $\sigma_{\delta p_{T}}$ rises with \sNN due to the increase 
in particle multiplicity at higher collision energies. These results are consistent with previous studies~\cite{Haake:2018hqn,Mengel:2023mnw} in that the 
deep neural network exhibits significantly improved momentum reconstruction resolution when compared to the area method. As expected, the shallow neural 
network achieves some of the increase obtained by the deep neural network over the area method but is unable to fully reproduce the momentum resolution. 
For \Au collisions with small jet resolution parameters, the multiplicity method matches the performance of the deep neural network and in \Pb collisions 
is able to out-perform the deep neural network at large jet momentum. The multiplicity method and shallow neural network method show increasingly similar 
jet momentum resolution to each other with increasing jet resolution parameter.

\begin{figure*}
  \centering
  \includegraphics[width=2\columnwidth]{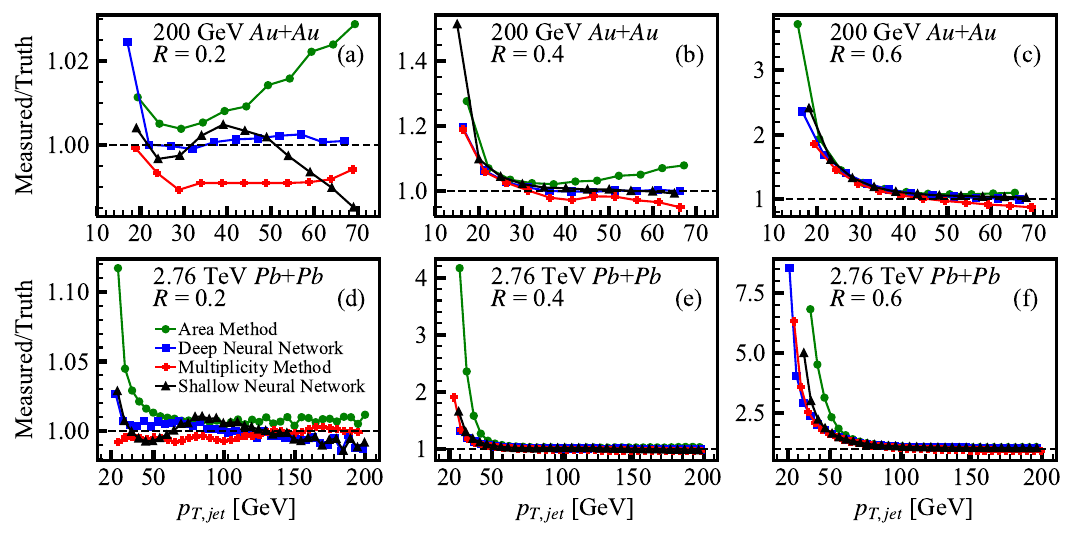}
  \caption{Ratio of the reconstructed jet spectrum over the truth spectrum for \Au collisions at \sNN = 200 GeV and \Pb collisions at \sNN = 2.76 
  TeV for jet resolution parameters $R=0.2, 0.4,$ and $0.6$. Low momentum points for the area method and shallow neural network methods for LHC energies at $R$ = 0.4 and $R$ = 0.6 are off scale.  
  }
  \label{fig:UncorrectedJetRatio}
\end{figure*}

 The fidelity of event-by-event subtraction of underlying event is not the sole metric which determines performance of a background subtraction method. 
 The ability of each method to suppress contributions from combinatorial jets in low jet \pT regions is demonstrated with the ratios of the reconstructed 
 jet spectra to the true jet spectra, shown in~\fref{fig:UncorrectedJetRatio}. These ratios show that the contributions from combinatorial jets decrease 
 with increasing jet momentum for all methods, with all jet resolution parameters, and for both collision energies. The difference between the area method 
 and the multiplicity, deep neural network, and shallow neural network methods can be seen at the lowest momentum bins where the area method is the 
 farthest from one for all jet samples. Contributions from combinatorial jets in low jet momentum regions determine where unfolding is stable. 
 
\begin{figure*}
  \centering
 \includegraphics[width=2\columnwidth]{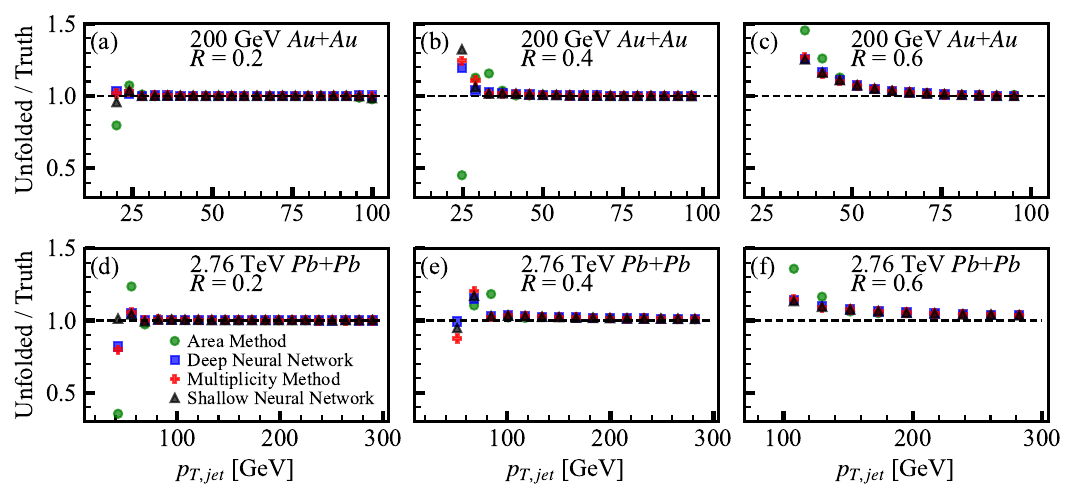}
  \caption{Ratios of unfolded jet spectrum over the truth jet spectrum for \Au collisions at \sNN = 200 GeV and \Pb collisions at \sNN = 2.76 TeV for 
  jet resolution parameters $R=0.2, 0.4,$ and $0.6$. Points that do not converge within $1.0 \pm 0.3$ are omitted.}
  \label{fig:UnfoldedJetRatio}
\end{figure*}

We find that the width of distribution alone does not significantly impact the uncertainties and that the kinematic reach of each method is 
driven by its ability to suppress combinatorial jets at low momentum. The ratios of the unfolded spectra to the true jet spectra are shown 
in~\fref{fig:UnfoldedJetRatio}. Each of shown ratios are after 5 iterations of Bayesian unfolding. Fluctuations away from one at lower jet momentum 
are where the given method becomes unstable due to overwhelming contributions from combinatorial background. The deep neural network, shallow neural 
network, and multiplicity methods are stable in unfolding to lower jet momenta than the area method. This indicates that experimental measurements can 
be extended to lower jet momenta using the deep neural network, shallow neural network, and multiplicity methods instead of the area method. The lower 
kinematic bound for jets in \Au collisions is extended at least 10 GeV for all jet parameters using methods other than the area method. Jets in \Pb collisions 
are extended down by 10-20 GeV for all jet resolution parameters. We find that the multiplicity method provides the same kinematic range extension of the deep neural network method. 

\begin{figure}
    \centering
    \includegraphics[width=\columnwidth]{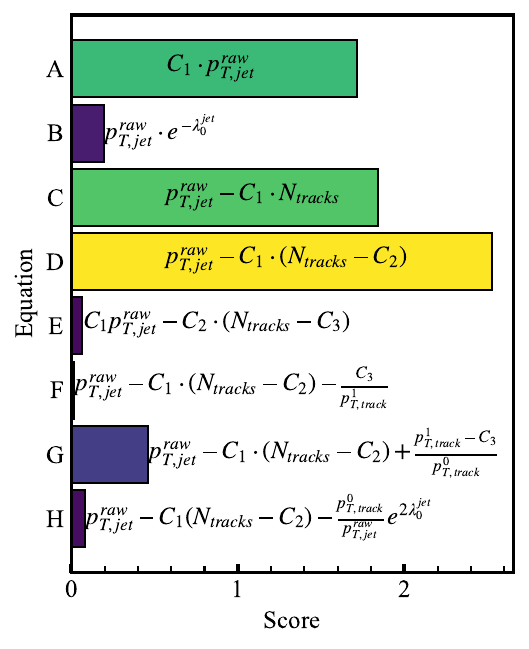}
    \caption{The eight highest scoring expressions from symbolic mapping of a deep neural network trained on R = 0.4 jets from \sNN = 200 
    GeV \Au collisions. The score is defined as the negative derivative of the loss function with respect to equation complexity (eq.~\ref{Eqn:Score}).}
    \label{fig:EquationScore}
\end{figure}

The similarities between the multiplicity method and the deep neural network method can be understood since the fluctuations in the background 
are well described by eq.~\ref{eq:DeltaPtwidths_flow}~\cite{TANNENBAUM200129,ALICE:2012nbx,Hughes:2020lmo}, where the multiplicity is the dominant 
variable in the standard deviation. This relationship is learned by the deep neural network~\cite{Mengel:2023mnw}.
We find that the functional representation of the mapping between the input jet features and the truth jet momentum learned by both the deep 
neural network and shallow neural network, is directly comparable to the multiplicity method. To demonstrate the output of PySR, the eight highest 
scoring expressions for the deep neural network trained on R = 0.4 jets from \sNN = 200 GeV \Au collisions are shown in~\fref{fig:EquationScore}. 
This mapping was repeated for the deep neural network and shallow neural network trained on jets samples with jet resolution parameter R = 0.2, 0.4, 
and 0.6 in both \sNN = 200 GeV \Au collisions and \sNN = 2.76 TeV \Pb collisions. 

\begin{figure}
  \centering
  \includegraphics[width=\columnwidth]{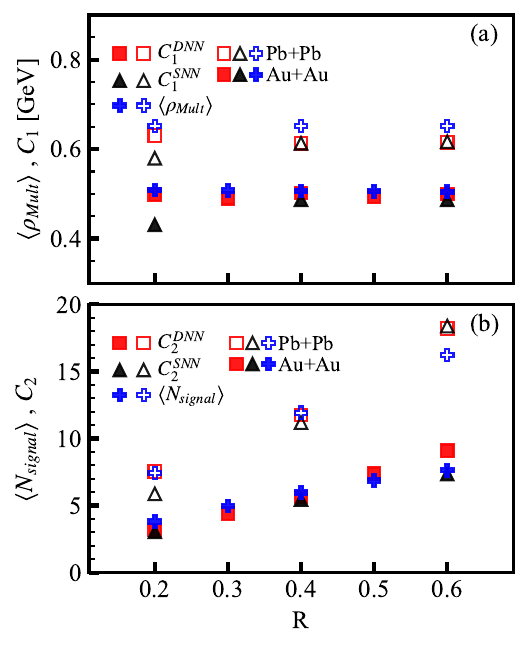}
  \caption{PySR optimization constants compared to average value of multiplicity method parameters versus jet resolution parameter for \Au 
  collisions at \sNN = 200 GeV and \Pb collisions at \sNN = 2.76 TeV for jet resolution parameters $R=0.2, 0.4,$ and $0.6$. 
  }
  \label{fig:pysr_params}
\end{figure}

In both networks, for all jet resolution parameters and both collision energies, the symbolic regression found that 
the best description of the deep neural network has the functional form
\begin{equation}
    p_{T,jet}^{Corr. PySR} = p_{T,jet}^{Uncorr.} - C_{1}\cdot (N_{tracks}- C_{2}) , \label{Eq:pTpysr}
\end{equation}
where the two parameters, $C_{1}$ and $C_{2}$, are optimization constants defined by PySR. The result of the symbolic regression implies 
that each of the neural networks relies heavily on input features such as the total number of particles in the jet $N_{tot}$ and the uncorrected 
jet momentum $p_{T,Jet}^{tot}$, both variables which are native to the multiplicity method.
These parameters are plotted in \fref{fig:pysr_params} and compared to the average value of the parameters used in the multiplicity method. 
We find that the symbolic regression parameters $C_{1}$ and $C_{2}$ for both the shallow and deep neural networks are comparable to the averages 
of those for the multiplicity method, $\langle \rho_{Mult} \rangle $ and $\langle N_{signal}\rangle $, respectively, with greater 
deviations at LHC energies and larger $R$. The PySr parameters from the shallow neural network mapping show larger discrepancies from 
those used in the multiplicity method at small jet resolution parameter and in \Pb collisions systems. Neither the shallow nor deep 
neural networks were given the average momentum of a background particle $\rho_{Mult}$ as input. Therefore, the neural network is able 
to learn the median momentum per particle in the event and the average number of signal particles in the jet during training. 
This indicates that the deep neural network and shallow neural networks are using a relationship similar to the multiplicity method to predict jet momenta. 

Using interpretable forms of machine learning, such as symbolic mapping of a neural network, allows up to make use of expert domain knowledge. 
The optimization parameters from PySR would otherwise not have a clear physical interpretation. Since these parameters are understood in the multiplicity method, 
it is possible to assign a physically motivated uncertainty to them.  Assumptions inherent in the method can then be understood.

%%%%%%%%%%%%%%%%%%%%%%%%%%%%%%%%%%%%%%%%%%%%%%%%%%%%%%%%%%%%%%%%%%%%%%%%%%%%
% Conclusion
%%%%%%%%%%%%%%%%%%%%%%%%%%%%%%%%%%%%%%%%%%%%%%%%%%%%%%%%%%%%%%%%%%%%%%%%%%%%
\section{Conclusion}

Our research reinforces the necessity of using interpretable machine learning techniques for scientific problems, particularly in the realm of 
high-energy physics. It is crucial to acknowledge the limitations of overly-complex machine learning methods that lack interpretability and 
require machine learning methods that satisfy strict interpretability criteria, outlined in our previous studies~\cite{Mengel:2023mnw}. 
Symbolic regression, yielding interpretable formulas, stands out as a promising approach that fulfills these criteria.

We demonstrate the over-complexity of a deep neural network to describe a relatively simple problem of jet background subtraction. 
The neural network pruning procedure described in~\ref{Sec:MachineLearning} provides a shallow neural network that not only enhances 
efficiency but also promotes transparency. This transparency allows for researchers to better assess the biases of a neural network 
from training data. We show that this network simplification procedure does not affect ability to for the network to learn the underlying 
relationship sufficiently jet multiplicity and background fluctuations~\cite{TANNENBAUM200129,ALICE:2012nbx,Hughes:2020lmo}. 
Discrepancies in performance of the shallow neural network, the deep neural network, and the multiplicity method seen in~\fref{fig:ReconstructionWidthvspT} 
are negligible after the unfolding process, indicating the methods all provide comparable performance.

This study builds upon our previous investigation into the relationship between  background and jet multiplicity~\cite{Mengel:2023mnw}. 
The demonstrated ability to increase kinematic range, across a wide range of jet resolution parameters, is further motivation for using the 
multiplicity method in the heavy ion jet measurements. This work provides a clear path of when and how to apply the multiplicity method with 
estimations of $N_{signal}$ from Monte Carlo~~\cite{ALICE:2014dla}, along with constraints of in medium modification using measured jet fragmentation 
functions~\cite{CMS:2014jjt,ATLAS:2018bvp,PHENIX:2020alr}. The convergence between the empirically derived multiplicity method and the formula 
generated through symbolic regression of both a deep and pruned network attests to  robustness of symbolic regression to pruning.

%%%%%%%%%%%%%%%%%%%%%%%%%%%%%%%%%%%%%%%%%%%%%%%%%%%%%%%%%%%%%%%%%%%%%%%%%%%%
% Acknowledgements
%%%%%%%%%%%%%%%%%%%%%%%%%%%%%%%%%%%%%%%%%%%%%%%%%%%%%%%%%%%%%%%%%%%%%%%%%%%%
\section{Acknowledgements}

We are grateful to Larry Lee and David Stewart for useful discussions and feedback on the manuscript. 
This work was supported in part by funding from the Division of Nuclear Physics of the U.S. Department of Energy 
under Grant No. DE-FG02-96ER40982. This work was performed on the computational resources at the Infrastructure for 
Scientific Applications and Advanced Computing (ISAAC) supported by the University of Tennessee.

%%%%%%%%%%%%%%%%%%%%%%%%%%%%%%%%%%%%%%%%%%%%%%%%%%%%%%%%%%%%%%%%%%%%%%%%%%%%
% appendix
%%%%%%%%%%%%%%%%%%%%%%%%%%%%%%%%%%%%%%%%%%%%%%%%%%%%%%%%%%%%%%%%%%%%%%%%%%%%
\newpage
\appendix
\section{Feature Scores}\label{Sec:FeatureScores}

\begin{figure*}
  \centering
  \includegraphics[width=\columnwidth]{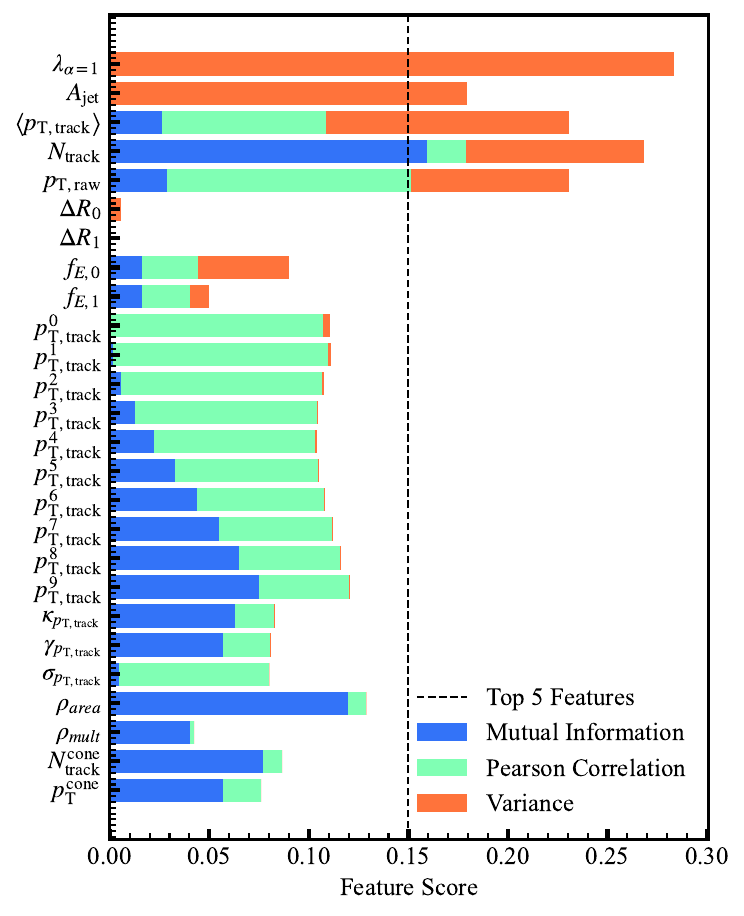}
  \caption{Total feature scores for available jet input features relating to $p_{T,jet}^{Truth}$. Each score is the sum of the variance, 
  mutual information, F-score, and Pearson Correlation Coefficient~\cite{features}.}
  \label{fig:FeatureScores}
\end{figure*}

Neural networks have shown great potential in achieving high-precision jet \pT resolution. The choice of input features 
significantly influences the performance of these networks~\cite{features}. This study explores the application of 
feature selection techniques, including mutual information, variance threshold, F-score, and Pearson correlations, 
to identify the most informative features for optimizing neural network-based jet \pT subtraction. Each of the 
aforementioned feature selection techniques offers unique insights into the relevance and information content of 
input features for jet \pT regression. These techniques assist in identifying the most informative and influential 
features, enabling the construction of more accurate models and potentially improving the overall performance of the 
regression task. The results of this feature optimization is shown in \fref{fig:FeatureScores}. 

While each feature selection technique provides valuable insights into the relevance of input features for jet \pT subtraction, 
relying on a single technique may have limitations. Therefore, an amalgamation of the scores obtained from multiple 
techniques can offer a more comprehensive perspective on feature importance. Combining the scores from different 
techniques allows us to take advantage of the unique perspectives and strengths of each method. For example, mutual 
information captures non-linear dependencies, while variance thresholding focuses on variability. The F-score assesses 
the overall impact of features, and Pearson correlations highlight linear relationships. By considering multiple techniques, 
we can identify features that consistently score high across different selection methods, indicating their 
strong relevance to the regression task~\cite{featurescores}.

\subsubsection{Variance Threshold}

The variance threshold technique focuses on the variability of features within the dataset. It sets a threshold on the variance value,
\begin{equation}
    \sigma^{2} = \sum_{i} \frac{(x_{i} - \mu)^{2}}{N}    
\end{equation}
and removes features that have a variance below this threshold. This technique is particularly useful when dealing 
with high-dimensional datasets, as it helps eliminate low-variance features that are likely to carry minimal 
information or exhibit little variability~\cite{features}. By removing such features, we reduce the dimensionality 
of the input space, simplifying the model and potentially improving the training and inference efficiency.

\subsubsection{Mutual Information}\label{Sec:MutualInfo}

Mutual information highlights features that exhibit a strong statistical dependence on jet pt, helping identify 
informative features with high predictive power. Mathematically the mutual information between two random variables $X$ and $Y$ is 
defined as the distance between the joint entropy distribution $H(x,y)$ and the sum of the individual entropy $H(x)$ and $H(y)$,
\begin{equation}
    I(x;y)\equiv H(x) + H(y) - H(x,y)
\end{equation}
\noindent where the  $H(n)$ is the Shannon entropy, $H(x) \equiv -\sum_{i} p(x_{i})\log (p(x_i))$, for variable $n$. 
The mutual information is applied to the process of feature selection but providing a measure of the relevance an input 
feature has in predicting the target variable~\cite{Vergara_2013}. The mutual information for all 29 available input features 
was calculated to the target $p_{T,jet}^{Truth}$ as a means of limiting the training input phase space to 
features with the most discriminatory power~\cite{PhysRevE.69.066138}.

\subsubsection{Pearson Correlation}

Pearson correlation coefficient measures the linear relationship between two variables. By calculating the correlation 
between each feature and the  $p_{T,jet}^{Truth}$ , we can identify features that exhibit strong linear associations 
with the target variable~\cite{pearson_vii_1895,features}. High absolute correlation values indicate features that are 
more likely to contribute to the regression model's predictive power. Pearson correlations provide insights into the 
strength and directionality of the relationships, helping to select features that have significant predictive capabilities. 
Mathematically the Pearson correlation coefficient between two variables is defined as
\begin{equation}
    \rho_{x,y} \equiv \frac{\mathbb{E}[(x-\mu_x)(y-\mu_y)]}{\sigma_x \sigma_y},
\end{equation}
\noindent where $\mathbb{E}$ is the expectation value, $\mu_n$ is the mean of variable $n$ and $\sigma_n$ is the standard deviation of variable $n$. 

%%%%%%%%%%%%%%%%%%%%%%%%%%%%%%%%%%%%%%%%%%%%%%%%%%%%%%%%%%%%%%%%%%%%%%%%%%%%
% References
%%%%%%%%%%%%%%%%%%%%%%%%%%%%%%%%%%%%%%%%%%%%%%%%%%%%%%%%%%%%%%%%%%%%%%%%%%%%
\bibliography{bib}% Produces the bibliography via BibTeX.

\end{document}